\pdfoutput=1
\documentclass[twocolumn,letterpaper]{IEEEAerospaceCLS}

\usepackage{graphicx}
\usepackage[colorlinks=true, urlcolor=blue, linkcolor=black, citecolor=black]{hyperref} 
\usepackage{textcomp}
\usepackage{xcolor}
\usepackage{subcaption}

\usepackage{booktabs} 
\usepackage{tabularx}  
\usepackage{array} 
\usepackage[acronym]{glossaries}
\newacronym{3gpp}{3GPP}{3rd Generation Partnership Project}
\newacronym{5g}{5G NR}{fifth generation new radio}
\newacronym{6g}{6G}{sixth generation}
\newacronym{5gc}{5GC}{5G core}
\newacronym{ai}{AI}{artificial intelligence}
\newacronym{a2a}{A2A}{air-to-air}
\newacronym{a2g}{A2G}{air-to-ground}
\newacronym{ar}{AR}{augmented reality}
\newacronym{api}{API}{application programming interface}
\newacronym{achem}{ACHEM}{AERPAW Channel Emulator}
\newacronym{aerpaw}{AERPAW}{Aerial Experimentation and Research Platform for Advanced Wireless}
\newacronym{av}{AV}{autonomous vehicle}
\newacronym{avn}{AVN}{autonomous vehicle network}
\newacronym{bs}{BS}{base station}
\newacronym{chem}{CHEM}{channel emulator}
\newacronym{cir}{CIR}{channel impulse response}
\newacronym{dac}{DAC}{digital-to-analog converter}
\newacronym{darpa}{DARPA}{Defense Advanced Research Projects Agency}
\newacronym{dl}{DL}{downlink}
\newacronym{doa}{DoA}{direction-of-arrival}
\newacronym{dt}{DT}{digital twin}
\newacronym{dtn}{DTN}{digital twin network}
\newacronym{enb}{eNB}{E-UTRAN NodeB}
\newacronym{epc}{EPC}{Evolved Packet Core}
\newacronym{fifo}{FIFO}{first-in-first-out}
\newacronym{fmcw}{FMCW}{frequency-modulated continuous wave}
\newacronym{fpga}{FPGA}{field programmable gate array}
\newacronym{gbsm}{GBSM}{geometry-based stochastic model}
\newacronym{gnb}{gNB}{Next-Generation Node B}
\newacronym{gnss-do}{GNSS-DO}{global navigation satellite system disciplined oscillator}
\newacronym{hitl}{HITL}{hardware-in-the-loop}
\newacronym{iot}{IoT}{Internet-of-Things}
\newacronym{ioe}{IoE}{Internet-of-Everything}
\newacronym{itu-t}{ITU-T}{ITU Telecommunication Standardization Sector}
\newacronym{lam}{LAM}{large AERPAW multicopter}
\newacronym{lte}{LTE}{Long-Term Evolution}
\newacronym{ng}{NextG}{next-generation}
\newacronym{ntn}{NTN}{non-terrestrial network}
\newacronym{mmtc}{mMTC}{massive machine-type communications}
\newacronym{otw}{OTW}{over-the-wire}
\newacronym{oran}{O-RAN}{open radio access network}
\newacronym{pci}{PCI}{physical cell identity}
\newacronym{pps}{PPS}{pulse-per-second}
\newacronym{prb}{PRB}{physical resource block}
\newacronym{pt}{PT}{physical twin}
\newacronym{ptv}{PtV}{Physical-to-Virtual}
\newacronym{rf}{RF}{Radio Frequency}
\newacronym{rw}{RW}{real-world}
\newacronym{rsrp}{RSRP}{reference signal received power}
\newacronym{rssi}{RSSI}{received signal strength index}
\newacronym{sdr}{SDR}{software-defined radio}
\newacronym{snr}{SNR}{signal-to-noise ratio}
\newacronym{sitl}{SITL}{software-in-the-loop}
\newacronym{vr}{VR}{virtual reality}
\newacronym{uav}{UAV}{uncrewed aerial vehicle}
\newacronym{ue}{UE}{user equipment}
\newacronym{ul}{UL}{Uplink}
\newacronym{ugv}{UGV}{uncrewed ground vehicle}
\newacronym{usrp}{USRP}{universal software radio peripheral}
\newacronym{vna}{VNA}{vector network analyzer}
\newacronym{vtp}{VtP}{Virtual-to-Physical}
\newacronym{xr}{XR}{extended reality} 
\newacronym{qos}{QoS}{quality-of-service}
\newacronym{oac}{OAC}{over-the-air computation}
\newacronym{tof}{ToF}{time-of-flight}
\newacronym{fl}{FL}{federated learning}

\begin{document}
\title{Air-to-Air Channel Characterization for UAV Communications at 3.4 GHz}

\author{
An\i l G\"{u}rses\\ 
Electrical and Computer Engineering\\
North Carolina State University\\
Raleigh, NC 27606\\
agurses@ncsu.edu
\and 
John Kesler\\
Electrical and Computer Engineering\\
North Carolina State University\\
Raleigh, NC 27606\\
jckesle2@ncsu.edu
\and 
Mihail L. Sichitiu\\
Electrical and Computer Engineering\\
North Carolina State University\\
Raleigh, NC 27606\\
mlsichit@ncsu.edu
}

\maketitle

\thispagestyle{plain}
\pagestyle{plain}

\begin{abstract}
The proliferation of Uncrewed Aerial Vehicles (UAVs) in applications such as flying ad-hoc networks (FANETs), precision agriculture, disaster response, and future 6G integrated networks necessitates the development of accurate and robust Air-to-Air (A2A) wireless communication systems.
While existing research has predominantly focused on Air-to-Ground (A2G) links, the A2A channel remains significantly under-characterized, especially in the sub-6 GHz frequency bands critical for reliable data exchange.
Current A2A models often oversimplify the channel, relying on static assumptions that neglect the profound impact of the UAVs' three-dimensional mobility and the physical characteristics of the aerial platforms themselves.
This paper addresses this research gap by presenting a preliminary set of measurements for the 3.4 GHz A2A channel. We have developed a lightweight, reconfigurable, open-source channel sounder using USRP B210 Software-Defined Radios (SDRs) and a high-precision Global Navigation Satellite System-disciplined oscillator (GNSS-DO), deployed on two UAVs.
We conducted a measurement campaign at the Aerial Experimentation and Research Platform for Advanced Wireless (AERPAW) Lake Wheeler testbed, an ideal, instrumented rural environment for UAV experimentation.
The campaign featured a spherical flight trajectory around the second drone, designed to capture the dynamic channel characteristics during maneuvers, including circular orbits, various altitudes, and elevation angles.
From these data, we present a thorough analysis of the fundamental channel characteristics.
We extract and model the fading parameters from the channel measurements, including channel impulse response (CIR), and analyze their dependence on link geometry.
We also characterize the fading statistics, providing insights into the RMS delay spread for A2A links in this environment.
This foundational channel measurement dataset provides a more realistic and validated tool for the design, development, emulation, and performance evaluation of physical and MAC layer protocols for next-generation UAV communication networks.

\end{abstract} 

\tableofcontents

\section{Introduction}
As we enter an era where \glspl*{uav} are becoming ubiquitous and integral to global connectivity, the need for reliable and efficient communication between these aerial platforms is more critical than ever \cite{UAV_future_directions}.
\glspl*{uav} have seen a surge in applications ranging from precision agriculture \cite{PrecisionAgriUAV}, and disaster response \cite{DisasterResponseUAV}, to their envisioned role in future 6G integrated networks \cite{6g_UAV}.
Central to the success of these applications is the ability of \glspl*{uav} to communicate effectively with each other, forming robust \gls*{a2a} and \gls*{a2g} links. However, despite the importance of \gls*{a2a} communication, there remains a gap in our understanding of the \gls*{a2a} wireless channel, particularly in the sub-6 GHz frequency bands where both ends of the link are airborne.
The existing literature has predominantly focused on \gls*{a2g} links, leaving the \gls*{a2a} channel under-characterized \cite{survey_a2g_model}.

Current \gls*{a2a} channel models often rely on oversimplified assumptions, such as static environments or two-dimensional mobility patterns, which fail to capture the complex dynamics introduced by the three-dimensional movement of \glspl*{uav} and the physical characteristics of the aerial platforms themselves \cite{3GPP38901}.
These simplifications can lead to inaccurate predictions of link performance, ultimately hindering the design and deployment of effective \gls*{uav} communication systems.
Previous studies have been limited both in scope and scale, often focusing on specific scenarios or environments that do not generalize well to the diverse conditions \glspl*{uav} may encounter.
Most notably, there is a lack of comprehensive measurement campaigns that systematically explore the \gls*{a2a} channel characteristics across various altitudes, distances, and flight patterns.
This paper aims to bridge this research gap by presenting a comprehensive measurement for the 3.4 GHz \gls*{a2a} channel.

Although the variations in the \gls*{a2a} channel characteristics are expected to be less severe than those in \gls*{a2g} channels due to the reduced likelihood of ground-based obstructions, the dynamic nature of \glspl*{uav} introduces unique challenges that must be addressed.
These undefined and oversimplified assumptions in \gls*{a2a} channel modeling can lead to significant discrepancies between predicted and actual performance, particularly in scenarios involving high mobility or complex maneuvers.
To accurately characterize the \gls*{a2a} channel, it is essential to conduct extensive measurement campaigns that capture the channel's behavior under various conditions.

To this end, we have used a lightweight, reconfigurable, open-source channel sounder using USRP B210 \glspl*{sdr} and a high-precision \gls*{gnss-do}, deployed on two \glspl*{uav} \cite{gurses2024a2g}.
The measurement was conducted at the \gls*{aerpaw} Lake Wheeler testbed \cite{aerpaw_website}, which provides an ideal, instrumented rural environment for \gls*{uav} experimentation.
The measurement featured a spherical flight trajectory around the second (center) drone designed to capture the dynamic channel characteristics during maneuvers for the spherical flight pattern that we have defined for this work.

In this study, we present a thorough analysis of the fundamental channel characteristics derived from the collected channel measurements.
We extract \gls*{cir} and model the fading parameters, analyzing their dependence on link geometry.
We also characterize the fading statistics, providing insights into the RMS delay spread for \gls*{a2a} links in this environment.
This foundational channel measurement dataset provides a more realistic and validated tool for the design, development, emulation, and performance evaluation of physical layer protocols and algorithms for \gls*{a2a} communications.

\section{Background and System Overview}
In this section, we provide a brief overview of the existing literature on \gls*{a2g} and \gls*{a2a} channel modeling and the design and implementation of our measurement system.

\subsection{Related Work}
Wireless channel modeling is a well-established field, with extensive research conducted on various types of channels, including rural, urban, and indoor environments \cite{uav_channel_sounding_survey,matolak_a2g_channel_1,matolak_a2g_channel_2,matolak_a2g_channel_3}. 
However, the unique characteristics of \gls*{a2a} channels, particularly involving \glspl*{uav}, have not been as thoroughly explored.
Most existing models for \gls*{a2a} channels are adaptations of terrestrial models, which may not accurately reflect the dynamics of aerial communication \cite{3GPP38901}.
Some studies have attempted to characterize \gls*{a2a} channels through simulations or limited measurements, but these efforts often lack the comprehensive scope needed to develop robust models \cite{a2a_channel_modeling_uav,rt_air_to_air}.
Recent advancements in \gls*{sdr} technology have enabled more flexible and cost-effective channel sounding systems, allowing for more extensive measurement campaigns.

\subsection{System Design and Implementation}
\begin{figure}[htbp]
\centering 
\includegraphics[width=\linewidth]{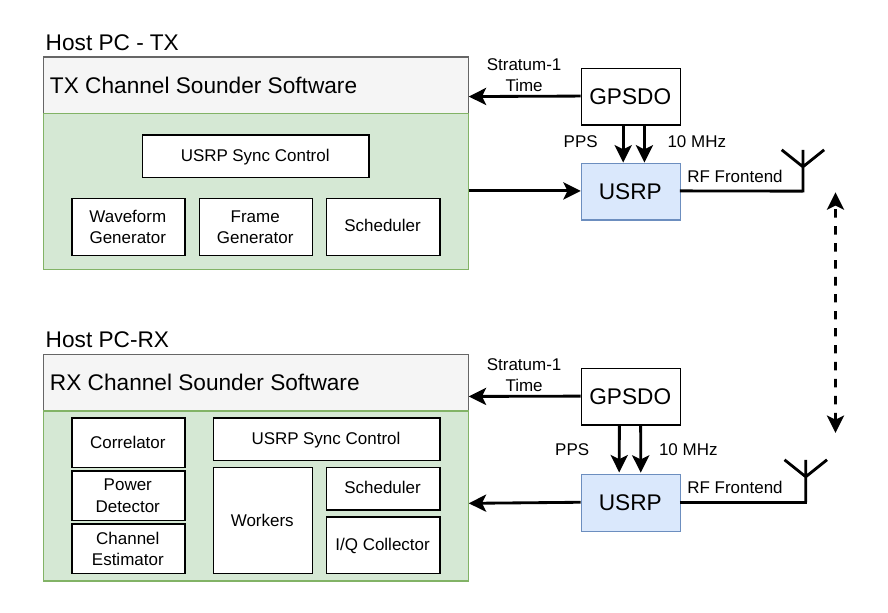}
\caption{Used channel sounder system block diagram.}
\label{UAV_Channel_Sounder}
\end{figure}
The channel sounder system is designed to be lightweight and reconfigurable, making it suitable for deployment on \glspl*{uav}. 
The system consists of two main components: the transmitter and the receiver, both based on USRP \glspl*{sdr}.
The transmitter is responsible for generating and transmitting the probe signal, while the receiver captures the received signal for analysis.
The probe signal is configurable and can be adjusted based on the specific requirements of the measurement campaign.
A high-precision \gls*{gnss-do} is used to provide accurate timing and frequency synchronization between the transmitter and receiver, which is crucial for accurate channel measurements.
The \gls*{gnss-do} and the \gls*{uav} are fed with local RTK corrections to achieve centimeter-level positioning accuracy and maintain a consistent clock synchronization.
The system is controlled using a custom software interface that allows for configuration and real-time monitoring of the measurement process.
The block diagram of the channel sounder system is shown in Figure \ref{UAV_Channel_Sounder}.
More details about the hardware and software design can be found in \cite{gurses2024a2g}.

Vehicle control and flight path planning are managed using custom vehicle software, which provides precise control over the \glspl*{uav}'s trajectory and orientation during the measurement campaign.
The \glspl*{uav} are equipped with GPS modules to provide real-time position data, which is logged alongside the channel measurements for post-processing and analysis.

The channel sounding process is based on correlation, where the received signal is correlated with the known transmitted probe signal to extract the channel impulse response.
The raw I/Q samples are processed and stored for further analysis, including the extraction of \gls*{cir} and path loss. 
To ensure the integrity of the measurements, the USRPs are calibrated with a power meter before deployment, and the antennas are previously characterized to account for their gain and radiation patterns \cite{ant_measurement}.

The wireless channel is represented by its \gls*{cir}, which describes how the transmitted signal is altered by the channel.
The \gls*{cir} can be expressed as:
\begin{equation}
h(t, \tau) = \sum_{i=1}^{N} a_i(t) \delta(\tau - \tau_i(t))
\end{equation}
where \(a_i(t)\) and \(\tau_i(t)\) are the amplitude and delay of the \(i\)-th multipath component, respectively, \(N\) is the total number of multipath components, and \(\delta(\cdot)\) is the Dirac delta function.

The path loss \(PL(d)\) has been modeled for \gls*{a2a} channels using the log-distance path loss model:
\begin{equation}
PL(d) = PL(d_0) + 10\gamma \log_{10}\left(\frac{d}{d_0}\right) + X_\sigma
\end{equation}
where \(PL(d_0)\) is the path loss at a reference distance \(d_0\), \(\gamma\) is the path loss exponent, \(d\) is the distance between the transmitter and receiver, and \(X_\sigma\) is a Gaussian random variable representing shadow fading with standard deviation \(\sigma\).

Based on the extracted \gls*{cir}, we can compute various channel metrics, including the RMS delay spread, which provides insights into the characteristics of the channel.
RMS delay spread \(\tau_{rms}\) is calculated as:
\begin{equation}
\tau_{rms} = \sqrt{\frac{\sum_{i=1}^{N} P_i (\tau_i - \bar{\tau})^2}{\sum_{i=1}^{N} P_i}}
\end{equation}
where \(P_i\) is the power of the \(i\)-th multipath component, and \(\bar{\tau}\) is the mean delay.

\section{Experiment Setup and Methodology}
In this section, we describe the details of the measurement campaign, including the experimental setup and flight pattern.
\subsection{Measurement Campaign}

\begin{figure}[htbp]
\centering 
\includegraphics[width=\linewidth]{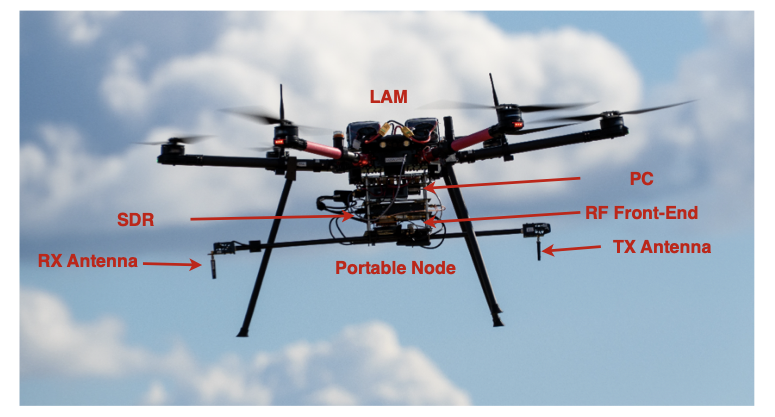}
\caption{LAM with portable node mounted during landing.}
\label{fig:LAM_w_PN}
\end{figure}

\begin{table}[h!]
    \centering
    \renewcommand{\arraystretch}{1.2}
    \resizebox{\linewidth}{!}{
    \begin{tabular}{l l}
        \hline
        \textbf{Parameter} & \textbf{Value} \\
        \hline
        Center Frequency              & 3400 MHz \\
        Sampling Rate                 & 56 MHz \\
        Transmit Power                & 19 dBm \\
        Transmitted Waveform          & Zadoff-Chu Sequence \\
        Sequence Length               & 2048 \\
        Root index                    & 89 \\
        Repetition of Sequence        & 4 \\
        Measurement Frequency         & 10 Hz \\
        Altitude (above ground level for center drone) & 65 m \\
        Flight Speed                  & 1.5 m/s \\
        Sphere Turns                  & 8 \\
        Sphere Radius                 & 20 m \\
        Ceiling Altitude              & 85 m \\
        Floor Altitude                & 45 m \\
        \hline
    \end{tabular}}
    \caption{Measurement Parameters}
    \label{tab:meas_parameters}
\end{table}

\begin{figure} [htbp]
\centering
\includegraphics[width=\linewidth]{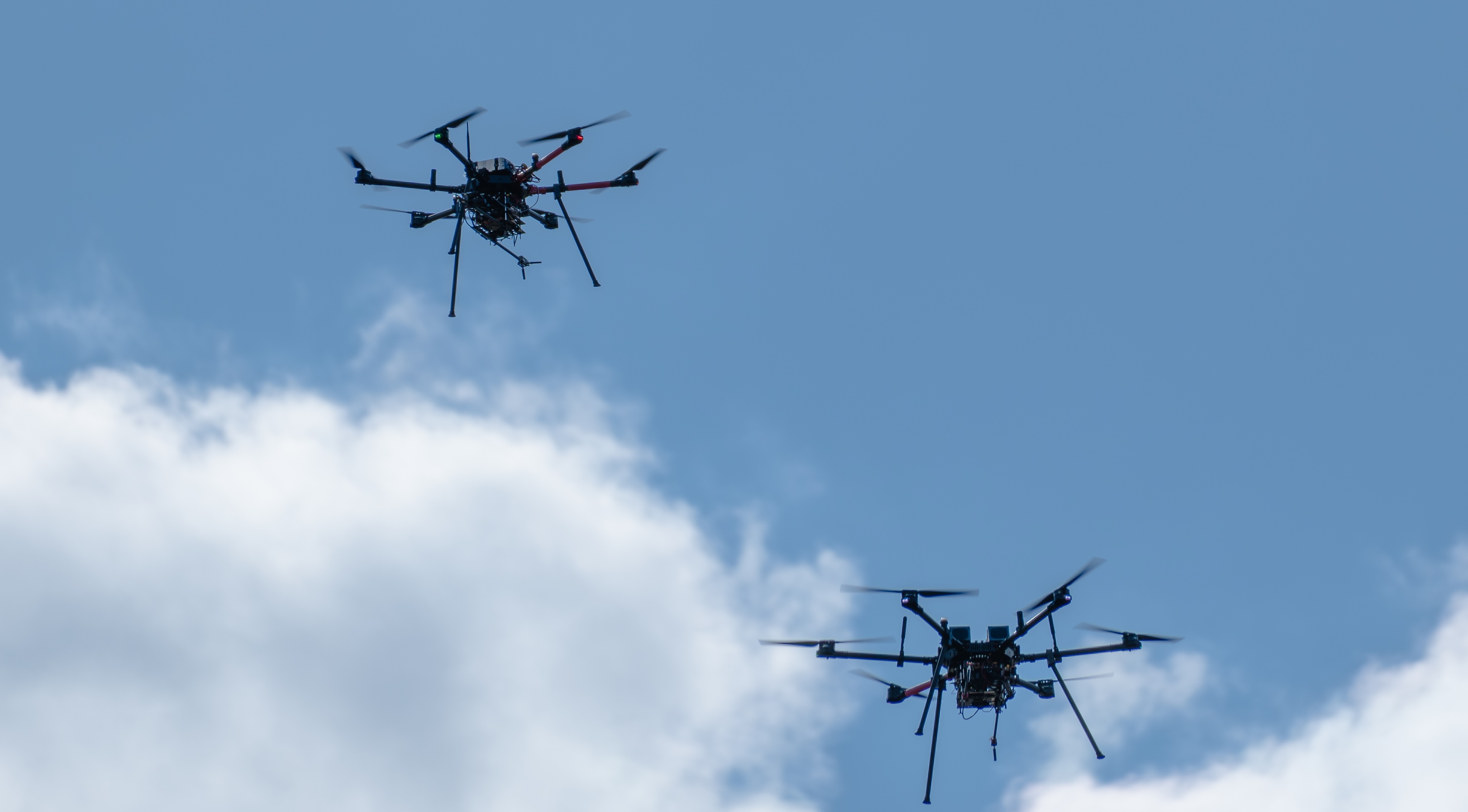}
\caption{UAVs with portable nodes mid flight.} 
\label{fig:Two_UAVs_in_Flight}
\end{figure}

The measurement campaign was conducted at the AERPAW Lake Wheeler testbed Raleigh, North Carolina, an instrumented environment ideal for repeatable UAV experiments.
The campaign involved two \glspl*{uav}, also identified as \gls*{lam}, one acting as the transmitter and the other as the receiver, as shown in Figure \ref{fig:Two_UAVs_in_Flight}.
Both \glspl*{uav} were equipped with the portable node shown in Figure \ref{fig:LAM_w_PN}, which includes the USRP \glspl*{sdr}, antennas, and RF front-end.
The transmitter \gls*{uav} was positioned at a fixed location at an altitude of 65m, while the receiver \gls*{uav} followed a predefined sphere flight trajectory around the transmitter.
The sphere trajectory was designed to capture a wide range of link distances, altitudes, and elevation angles, ensuring a comprehensive dataset for analysis.
The key measurement parameters are summarized in Table \ref{tab:meas_parameters}. 

The transmitter \gls*{uav} maintained a constant altitude of 65 meters above ground level, while the receiver \gls*{uav} varied its altitude between 45 and 85 meters, creating a spherical measurement pattern with a radius of 20m around the transmitter.
The flight speed of the receiver \gls*{uav} was kept at approximately 1.5 m/s to ensure stable measurements while capturing the dynamic nature of the \gls*{a2a} channel.
The measurement frequency was set to 10 Hz, allowing for dense sampling of the channel characteristics during the flight.

The probe signal used for the measurements was a Zadoff-Chu sequence, known for its good correlation properties and constant amplitude, which is ideal for channel sounding applications.
The sequence length was set to 2048 samples (chosen for FFT efficiency, despite the non-prime length slightly degrading the correlation properties), with a root index of 89, and the sequence was repeated four times to improve the signal-to-noise ratio (SNR) of the measurements and ensure reliable extraction of the \gls*{cir}.

\begin{figure} [htbp]
\centering
\includegraphics[width=\linewidth]{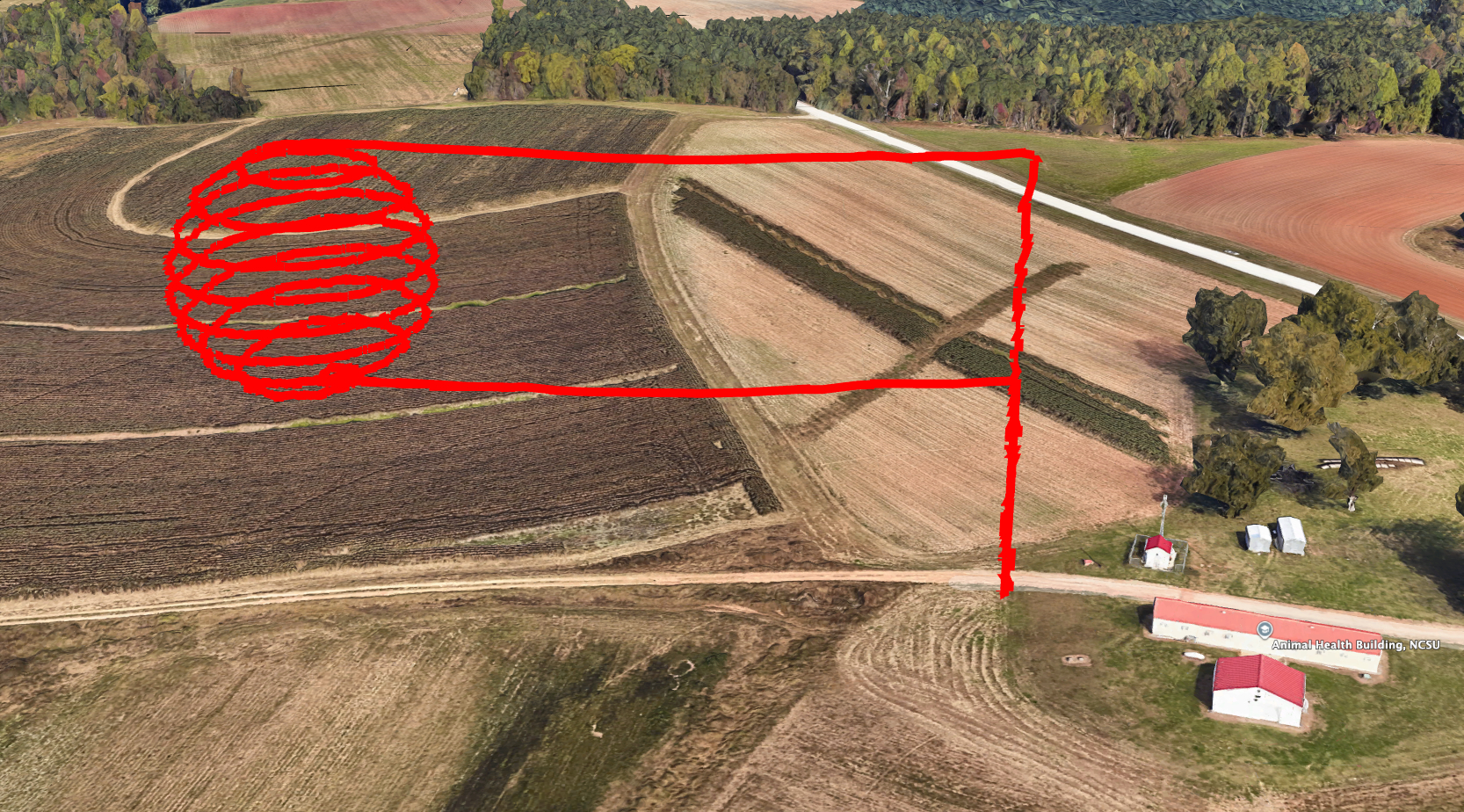}
\caption{Flight plan showing the sphere trajectory around the transmitter UAV.}
\label{fig:flight_plan}
\end{figure}

\subsection{Flight Pattern}
To evenly distribute the channel sounding samples in space around the transmitter \gls*{uav}, vehicle control software was written to fly the receiver \gls*{uav} along parametrically defined trajectories.
During data collection flights, a spherical trajectory was flown with the parameters defined in Table \ref{tab:meas_parameters}.
The path followed can be seen in Figure \ref{fig:flight_plan}. 
The path taken was described in the North-East-Up coordinate frame, as a function of time ($t$), radius ($R$), sphere center altitude ($Y$), number of turns ($n$), and path velocity ($v$) \gls*{uav}:

\begin{equation}
    U(t) = \left(\frac{v}{R}t\right)+Y-R
\end{equation}
\begin{equation}
    N(t) = R \sqrt{1-\left(\frac{U(t)-Y}{R}\right)^2}\cos\left(n\pi\frac{U(t)-Y}{R}\right)
\end{equation}
\begin{equation}
    E(t) = R \sqrt{1-\left(\frac{U(t)-Y}{R}\right)^2}\sin\left(n\pi\frac{U(t)-Y}{R}\right)
\end{equation}

Along the path, the receiver \gls*{uav}'s attitude was controlled so that it always pointed towards the transmitter, while the transmitter \gls*{uav} faced north.
This trajectory ensures that the receiver traverses a wide range of solid angles relative to the transmitter, providing a complete 3D characterization of the link.
Consequently, this approach decouples the impact of distance-based path loss from the orientation-dependent antenna gain and fuselage shadowing.

To maintain a consistent velocity along parametrically defined paths with arbitrary curvature, a non-linear trajectory-following control algorithm was used \cite{l1_control_algorithm}.
It was found that the \gls*{uav}s' default PID-based controllers were unable to minimize position error along trajectories that required high acceleration, leading to distortion of and oscillations along the trajectory.
The new controller's parameters were empirically tuned using a simulation to sufficiently damp the system for maneuvers of up to 10 $\frac{m}{s^2}$ of acceleration.

\section{Results and Analysis}

\begin{figure}[htbp]
\centering 
\includegraphics[width=\linewidth]{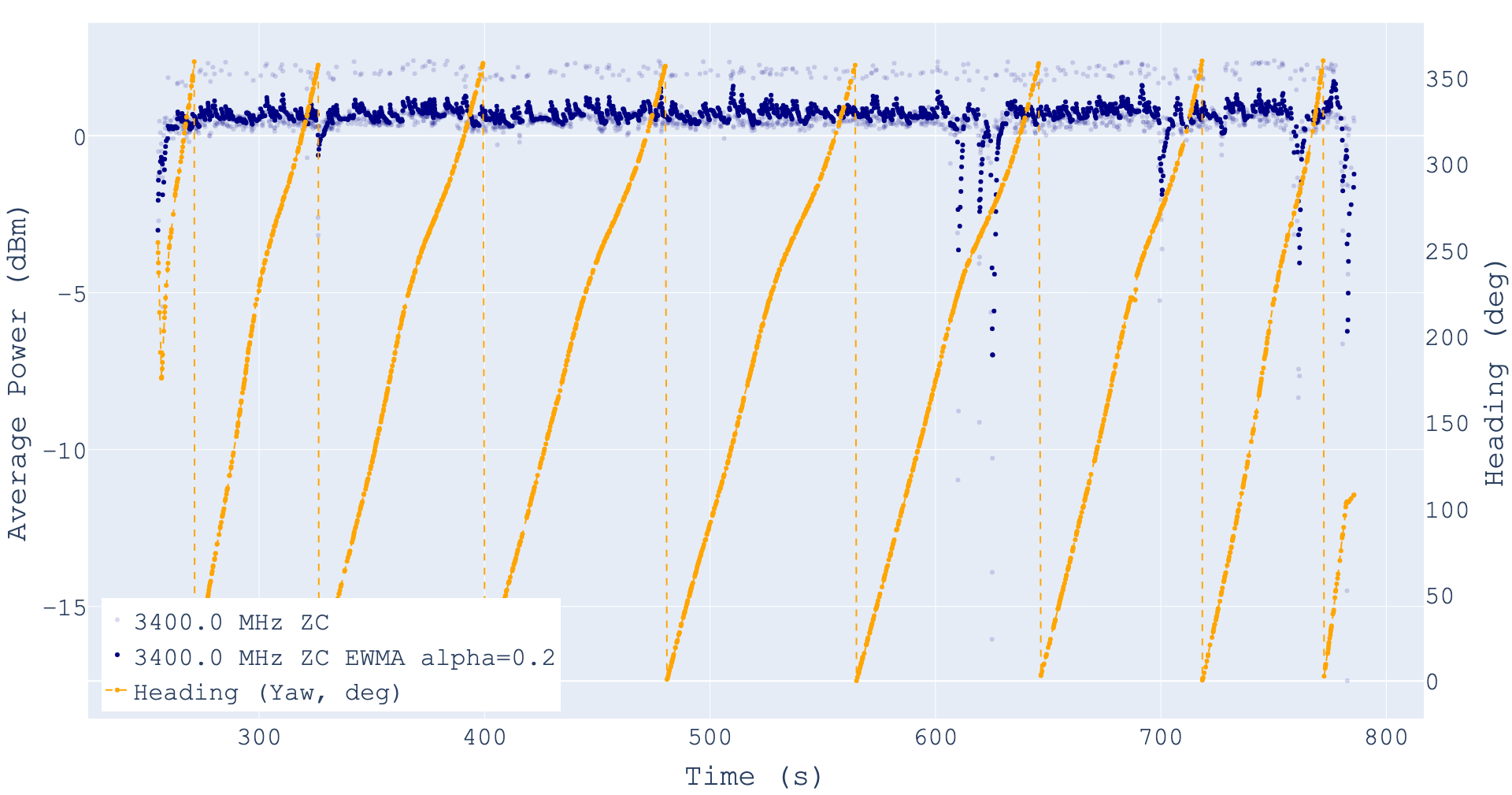}
\caption{Average received power as a function of altitude including heading information.}
\label{fig:Avg_PWR_vs_Alt}
\end{figure}
\begin{figure}[htbp]
\centering
\includegraphics[width=\linewidth]{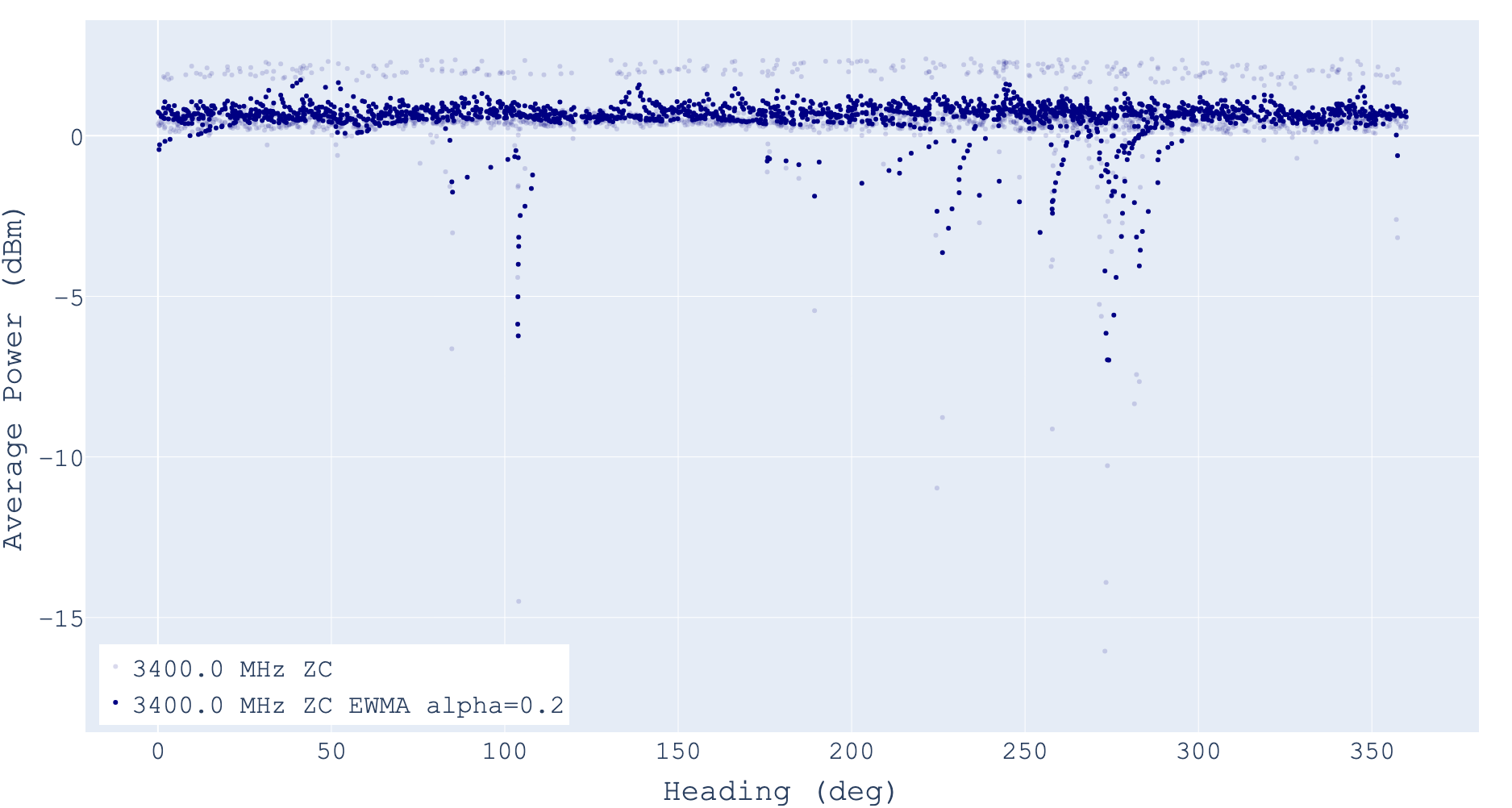}    
\caption{Average received power as a function of heading.}
\label{fig:Avg_power_vs_heading}
\end{figure}

\subsection{Large-Scale Fading Analysis}

Figure \ref{fig:Avg_PWR_vs_Alt} and \ref{fig:Avg_power_vs_heading} show the average received power as a function of altitude and heading, respectively, highlighting the influence of elevation angle and antenna orientation on signal strength.
The plot includes heading information, which provides insights into the directional characteristics of the antennas used in the measurement campaign.
The plot reveals a general trend of decreasing received power with increasing altitude on certain headings, which can be attributed to the antenna radiation patterns and the relative orientation of the \glspl*{uav}.
Deep fades are observed at specific altitudes and headings, likely due to the antenna nulls and the \gls*{uav} frame obstructing the line-of-sight path.
This behavior is also visible in the 3D power plot in Figure \ref{fig:Power_3D_Plot}.
When the available antenna measurement data is considered \cite{ant_measurement}, the observed power variations align with the expected antenna gain patterns, confirming the significant impact of antenna orientation on the received signal strength.
Although this is an expected behavior, it is important to note that the \gls*{a2a} channel is subject to other factors such as frame obstruction, multipath effects, and environmental influences, which can further complicate the received power patterns.
In particular, the conductive components of the UAV airframe, such as batteries and motors, can cause significant shadowing when positioned between the antennas.
Therefore, relying solely on free-space antenna patterns is insufficient for mission-critical A2A link planning.

\subsection{Small-Scale Fading Analysis}
\begin{figure}[htp]
\centering 
\includegraphics[width=\linewidth]{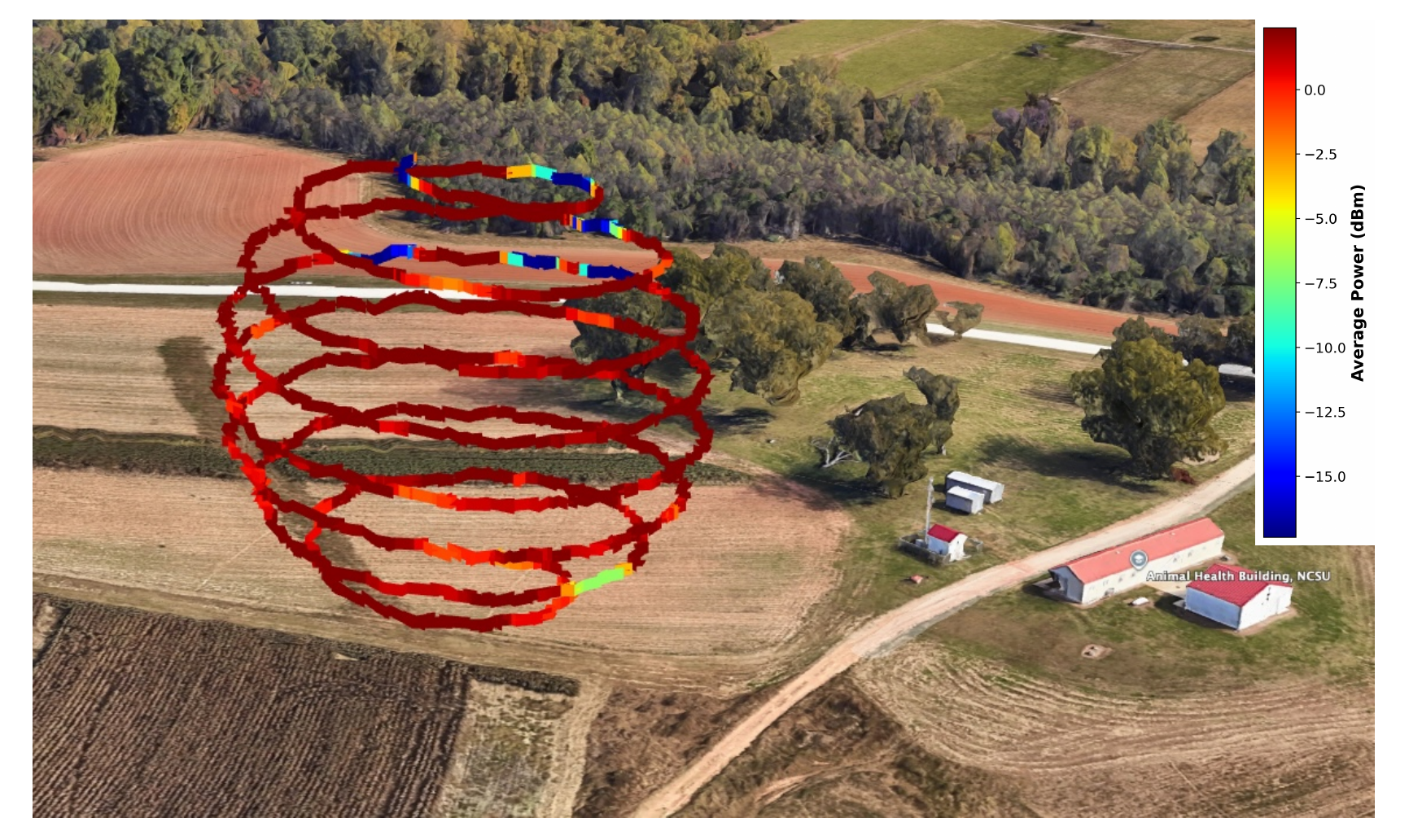}
\caption{Received power during the flight in 3D space.}
\label{fig:Power_3D_Plot}
\end{figure}

\begin{figure} [htbp]
\centering
\includegraphics[width=\linewidth]{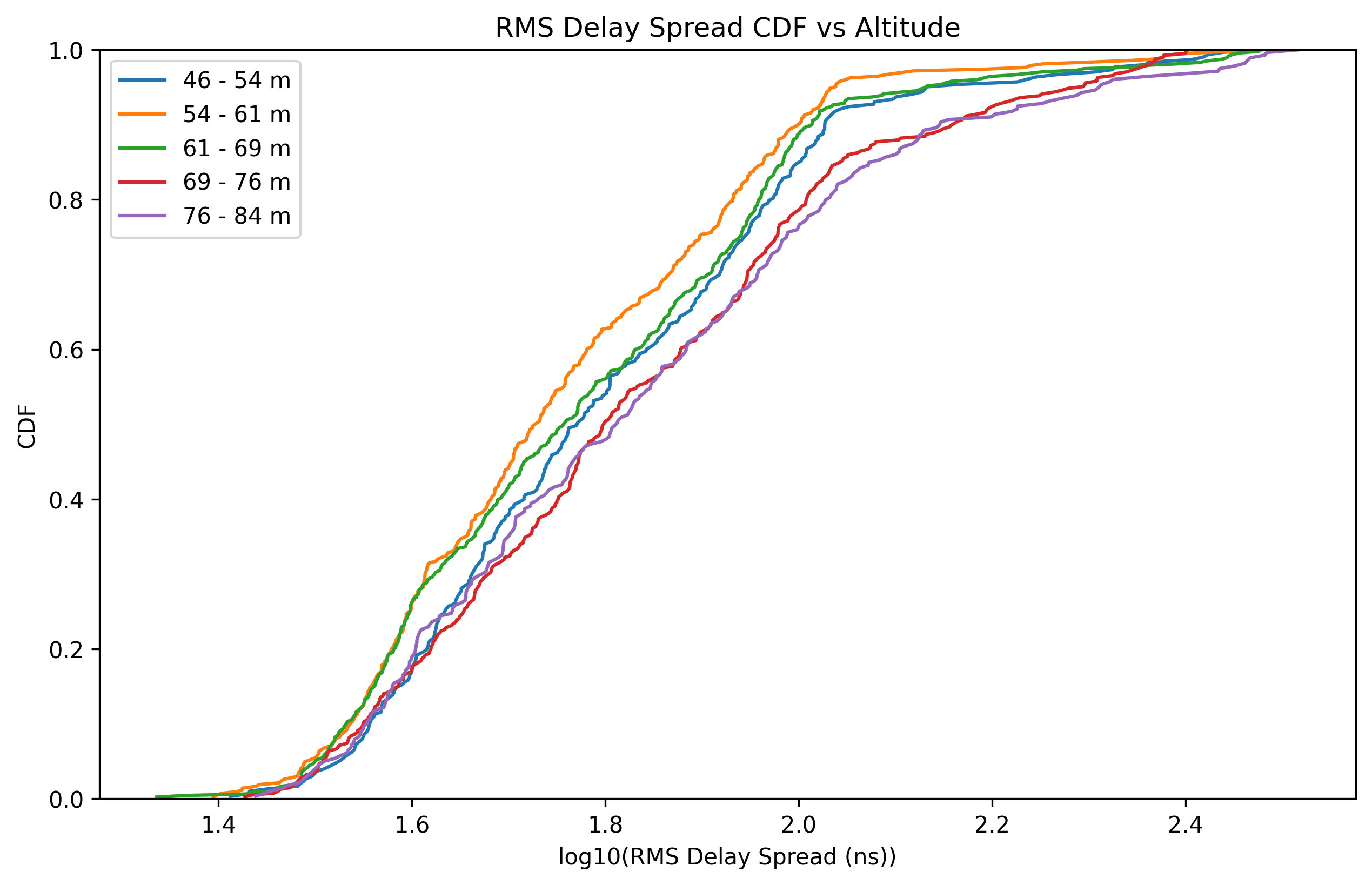}
\caption{CDF of RMS Delay Spread at Different Altitudes.}
\label{fig:rms_delay_spread_cdf_vs_altitude}
\end{figure}

The small-scale fading characteristics were analyzed by examining the \gls*{cir} and RMS delay spread.
The RMS delay spread was calculated from the \gls*{cir}, yielding values varying between 1.4 ns and 2.4 ns across all measurements.
Although the \gls*{a2a} channel is expected to exhibit minimal multipath effects, the presence of ground reflections and other environmental factors contributed to the observed delay spread.
Figure \ref{fig:rms_delay_spread_cdf_vs_altitude} shows the cumulative distribution function (CDF) of the RMS delay spread at different altitudes, indicating a slight increase in delay spread with altitude, likely due to the identifiable increased likelihood of ground reflections at higher altitudes.

\begin{figure}
\centering 
\includegraphics[width=\linewidth]{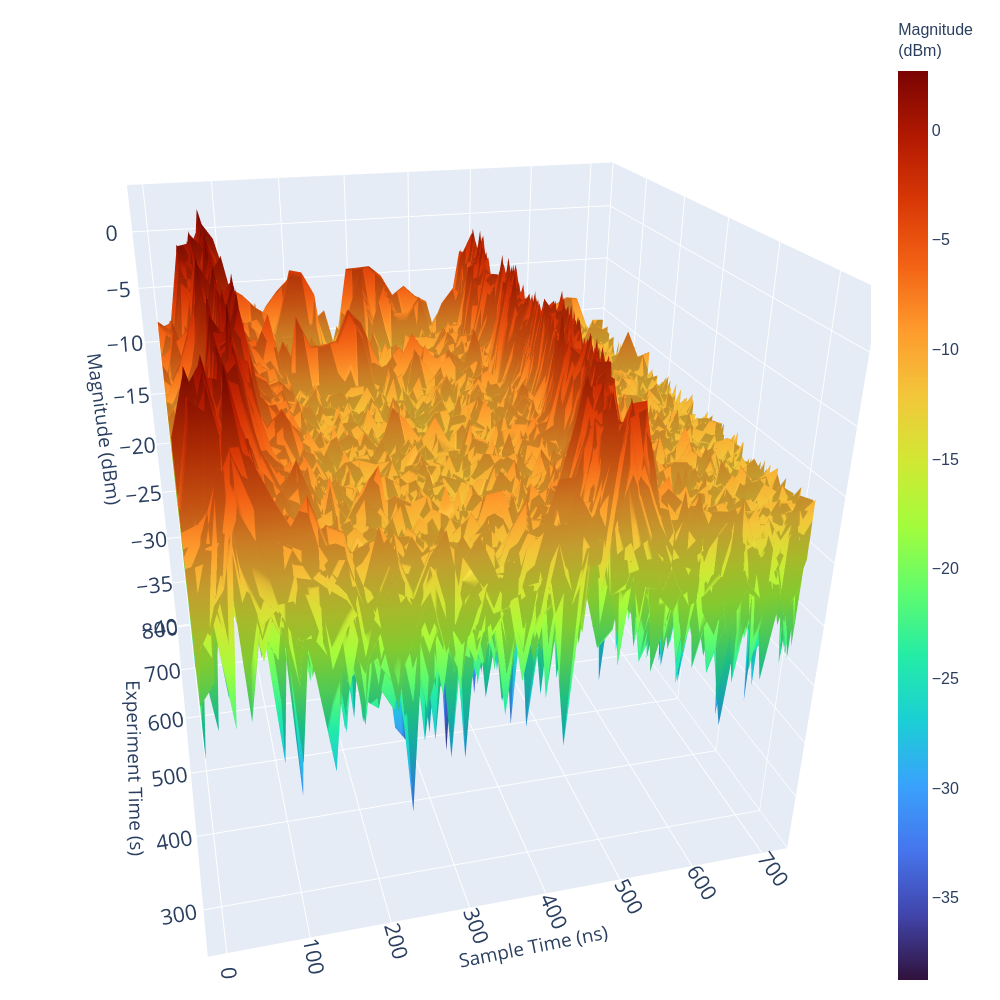}
\caption{Channel Impulse Response (CIR) showing multipath components during flight.}
\label{fig:CIR}
\end{figure}

Additionally, aforementioned reflections can be observed in the \gls*{cir} shown in Figure \ref{fig:CIR}, where multiple multipath components are visible, indicating the presence of reflections and scattering in the environment despite the aerial nature of the link.
This plot shows a snapshot of the \gls*{cir} during a segment of the flight, after the orbiter \gls*{uav} has reached the bottom of the sphere as a starting point and is ascending.
A detailed analysis of the multipath components revealed that the strongest path was typically the line-of-sight (LOS) component, with secondary paths corresponding to ground reflections and a strong reflection from 150 meters away, most likely from the nearby ground operations building as can be seen in Figure \ref{fig:flight_plan}.

\begin{figure*}[htbp]
\centering
\includegraphics[width=\linewidth]{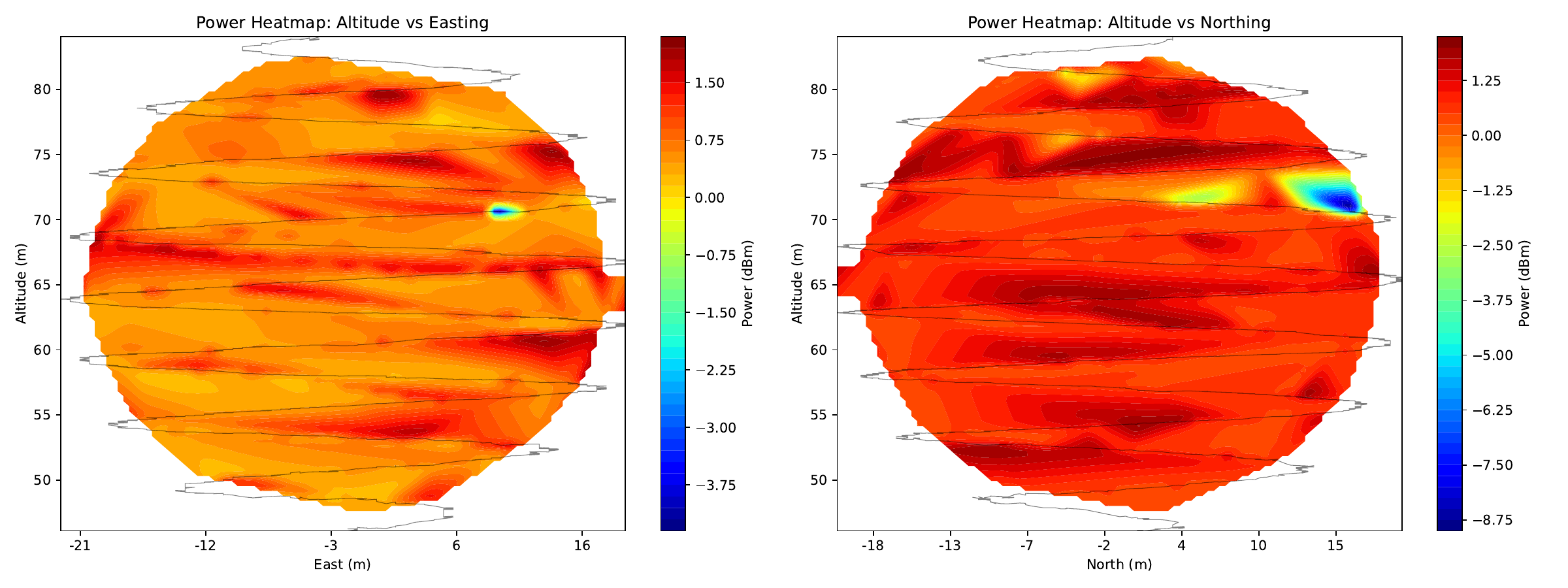}
\caption{Heatmap of Received Power as a function of East and North Coordinates at Different Altitudes.}
\label{fig:Power_Heatmap_Alt_vs_East_North}
\end{figure*}

Figure \ref{fig:Power_Heatmap_Alt_vs_East_North} illustrates the received power as a function of the eastings and northings coordinates at different altitudes.
The heatmaps reveal clear gradients in signal strength, with higher power levels concentrated in localized regions at mid-range altitude, where the receiver \gls*{uav} is closest to the transmitter \gls*{uav}.
Along the easting axis, power values gradually increase with altitude up to approximately 70 m before tapering off, whereas along the northing axis the distribution appears more irregular, with localized peaks between 60 m and 75 m.

Figure \ref{fig:Power_3D_Plot} provides a three-dimensional representation of the received power, highlighting the spatial distribution of signal strength around the center \gls*{uav}.
The plot clearly shows the spherical measurement pattern, with power levels varying based on the receiver's position relative to the transmitter.
The highest power levels are observed when the receiver's antenna is directly facing the transmitter without obstruction of the \gls*{uav}'s frame, showing the impact of the frame's position and orientation of the antennas.
Conversely, power levels decrease significantly as the receiver moves laterally away from the transmitter, suggesting the \gls*{uav} frame obstruction effects.

\section{Conclusion}
In this paper, we presented a comprehensive measurement and modeling campaign for the 3.4 GHz \gls*{a2a} channel using a lightweight, reconfigurable, open-source channel sounder deployed on two \glspl*{uav}.
The extensive measurement campaign conducted at the AERPAW Lake Wheeler testbed provided a statistically rich dataset, capturing the dynamic channel characteristics during various maneuvers on a unique spherical flight trajectory.
We analyzed the fundamental channel characteristics, extracting and modeling the fading parameters from the channel measurements.
We also analyzed the RMS delay spread, providing insights into the multipath characteristics.
The results highlighted the impact of antenna orientation and \gls*{uav} frame obstruction on the received signal strength, as well as the presence of multipath components despite the aerial nature of the link.
This foundational channel dataset has been published as open-source \cite{gurses_channel_sounder}.

\acknowledgements
This work was supported in part by the NSF PAWR Program under Grant CNS-193933. LLM was used to proofread and improve the grammar of this manuscript. The authors would like to thank the AERPAW team for their support during the measurement campaign.

\bibliographystyle{IEEEtran}
\bibliography{refs}

\end{document}